\documentclass[a4paper,fleqn,usenatbib]{mnras}

\usepackage{mathptmx}

\usepackage[T1]{fontenc}
\usepackage{ae,aecompl}

\usepackage{graphicx}	
\usepackage{amsmath}	
\usepackage{amssymb}	

\def\simlt{\mathrel{\hbox{\rlap{\hbox{\lower4pt\hbox{$\sim$}}}\hbox{$<$}}}}
\def\simgt{\mathrel{\hbox{\rlap{\hbox{\lower4pt\hbox{$\sim$}}}\hbox{$>$}}}}

\newcommand{\gw}{\textrm{GW150914}\,}

\newcommand{\tmerge}{t_{\textrm{merge}}}
\newcommand{\twr}{t_{\textrm{WR}}}
\newcommand{\twind}{t_{\textrm{wind}}}
\newcommand{\ttau}{t_{\tau}}

\newcommand{\xeq}{x_{\textrm{eq}}}

\newcommand{\async}{a_{\textrm{sync}}}

\newcommand{\chieff}{\chi_{\textrm{eff}}}
\newcommand{\veq}{v_{\textrm{eq}}}
\newcommand{\fb}{f_{\textrm{break}}}
\newcommand{\ftau}{f_{\tau}}

\title[Spin-based constraints on the merger time]{\gw: spin-based constraints on the merger time of the progenitor system}

\author[D. Kushnir et al.]{
Doron Kushnir,$^{1,2}$\thanks{E-mail: kushnir@ias.edu}
Matias Zaldarriaga,$^{1}$
Juna A. Kollmeier$^{1,3}$
and Roni Waldman$^{4}$
\\
$^{1}$School of Natural Sciences, Institute for
Advanced Study, Princeton, NJ, 08540, USA\\
$^{2}$John N.\ Bahcall Fellow\\
$^{3}$ Observatories of the Carnegie Institution of Washington,
  813 Santa Barbara Street, Pasadena, CA 91101, USA\\
$^{4}$Racah Institute of Physics, Hebrew University, Jerusalem, 9190401, Israel
}

\date{Accepted XXX. Received YYY; in original form ZZZ}

\pubyear{2015}

\begin{document}
\label{firstpage}
\pagerange{\pageref{firstpage}--\pageref{lastpage}}
\maketitle

\begin{abstract}
We explore the implications of the observed low spin of \gw within the context of stellar astrophysics and progenitor models.  We conclude that many of the recently proposed scenarios are in marked tension with this observation. We derive a simple model for the observed spin in the case that the progenitor system was a field binary composed of a black hole (BH) and a  Wolf--Rayet star and explore the implications of the observed spin for this model. The spin observation allows us to place a lower limit for the delay time between the formation of the BH+BH binary and the actual merger, $\tmerge$. We use typical values for these systems to derive $\tmerge\simgt10^{8}\,\textrm{yr}$, which proves to be an important diagnostic for different progenitor models. We anticipate the next series of events, and the associated spin parameters, will ultimately yield critical constraints on formation scenarios and on stellar parameters describing the late-stage evolution of massive stars.  
\end{abstract}




\section{Introduction}
\label{sec:Introduction}
The era of gravitational wave astronomy has arrived.   Both the strength of the signal in GW150914, as well as its form, were a striking revelation to all  \citep{Abbott2016PhRvL}.  In particular, the reported masses of the system  -- consistent with a black hole+black hole (BH+BH) binary merger with estimated BH masses\footnote{The error quotes both the $90\%$ credible interval and an estimate for the $90\%$ range of systematic error \citep{LIGO2016}.} of  $M_{1}=35.7^{+5.4\pm1.1}_{3.8\pm0.0}$, $M_{2}=29.1^{+3.8\pm0.2}_{4.4\pm0.5}$  --- are significantly `heavier' than  expected based on the known mass function of stellar mass BHs within the Milky Way.   The delightfully unexpected properties of GW150914 have unleashed the creative fury of the theoretical astrophysics community. As a result, there is an enormous range of proposed channels all aiming to produce BHs in the observed mass range.

In this article, we explore another property of GW150914 --- its observed (low) spin.  The effective inspiral spin parameter,
\begin{equation}\label{eq:chieff}
\chieff=\frac{M_{1}\vec{a}_{1}+M_{2}\vec{a}_{2}}{M_{1}+M_{2}}\cdot\hat{L}
\end{equation}
(where $\vec{a}_{1}$ and $\vec{a}_{2}$ are the dimensionless BH spins, $\vec{a}= c\vec{S}/GM^{2}$, and $\hat{L}$ is the direction of orbital angular momentum), is observed to be  $\chieff=-0.06^{+0.17\pm0.01}_{-0.18\pm0.07}$.

There are multiple proposed channels for the merger of stellar-mass BHs including, but not limited to isolated stellar field binary merger \citep[hereafter, the \textit{classical scenario},][]{Phinney1991,Tutukov1993,Belczynski2016}, dynamical formation in dense environments \cite[e.g. in globular clusters,][]{Sigurdsson1993}, and merger inside a massive star envelope \citep{Fryer2001,Reisswig2013,Loeb2016,Woosley2016}. Recently, an isolated stellar field binary scenario that includes chemically homogeneous stars has been discussed by \citet{Mandel2016,Marchant2016}, and \citet{dMink2016}. For a more complete reference list of proposed channels, see \citet{Abbott2016ApJL}. Some of these scenarios predict alignment of the BH spin and the orbital angular momentum. In this case $\chieff$ is just the mass-weighted mean of the dimensionless spins $a= c|\vec{S}|/GM^{2}$, which can generically be quite large in possible tension with the observed low $\chieff$. Following are the examples. 
\begin{enumerate}
\item  Formation inside a massive star envelope: the fission of the core to two BHs should generally lead to $a\sim1$ for each of them. For example, \citet{Reisswig2013} simulate the fission to two BHs within a common envelope and the subsequent BH--BH merger. In the final orbits before the merger, each simulated BH has $a\approx0.7$, which would lead to $\chieff\approx0.7$ compared to the low observed value $\chieff\ll 1$.
\item The chemically homogeneous stars scenario: \citet{Yoon2006} derived the condition $\veq\ge0.2\sqrt{GM/R}$, where $\veq=R\Omega$ is the equatorial velocity at the stellar radius, $R$, and $\Omega$ is the angular spin velocity of the star, as the threshold for homogenous evolution (see their fig. 3; larger $\veq$ are required for lower mass stars). Neglecting the effect of mass-loss during the collapse of a fully mixed star to a BH (which is justified later on), the dimensionless spin of the BH satisfies: 
\begin{eqnarray}\label{eq:M case}
a&\ge&0.2cr_{g}^{2}\sqrt{\frac{R}{GM}}\nonumber\\
&\approx&2.7\left(\frac{r_{g}^{2}}{0.075}\right)\left(\frac{R}{2\,R_{\odot}}\right)^{1/2}\left(\frac{M}{30\,M_{\odot}}\right)^{-1/2},
\end{eqnarray}
where $r_{g}^{2}$ is the (dimensionless) radius of gyration of the star related to the moment of inertia by $I=r_{g}^{2} R^{2}M$. 
Equation (\ref{eq:M case}) implies that both BHs should have close-to-maximal spins ($a\approx1$). This prediction of $\chieff\approx1$ is thus in seeming tension with the low observed value $\chieff\ll 1$. In a more recent study of homogenous evolution, \citet{Marchant2016} provide the threshold $a$ values for their models directly. All systems with $M_{1}+M_{2}\simlt100\,M_{\odot}$ that merge within Hubble time satisfy $a\simgt0.5$ for both components, leading to $\chieff\simgt0.5$, which also in tension with the low observed value $\chieff\ll 1$.
\item \citet{Perna2016} suggested that a disc forms around one of the BHs to later power a short gamma-ray burst. For the specific scenario that they proposed, the spin of this BH is $a\approx1$, which is also in tension with the low observed value $\chieff\ll 1$ for GW150914. Note, however, that by modifying the angular momentum profile of the BH progenitor, this scenario can be tuned to produce a BH with a low spin.
\end{enumerate}
These models, in their raw form, generally predict $\chieff$ close to unity.  The observation in GW150914 of $\chieff\ll 1$ thus already provides some constraining power on the details of these models. 

In this paper, we will analyse the \textit{classical scenario} and show that the spin observation allows us to place a lower limit on the delay time between the formation of the BH+BH binary and the actual merger, $\tmerge$. This arises because a short delay time corresponds to small separation between the primary BH and the star that evolves to become the second BH. For small separations, the primary BH applies torque to the star, spinning it towards synchronization. The angular momentum resulting from this process violates the observed upper limit given by the observed $\chieff$. We derive a simple model to describe this process, and we elucidate the basic considerations here so that future events may be similarly used to constrain models. We use typical values for these systems to derive for $\gw$ $\tmerge\simgt10^{8}\,\textrm{yr}$.

Our paper is organized as follows.  In Section~\ref{sec:a range} we consider the angular momentum evolution of the secondary star as it collapses to a BH, examining the competing effects of wind-losses and torque-gains.  In Section~\ref{sec:implications} we consider the allowed range of scenarios and demonstrate the application of our limits.    In Section~\ref{sec:caveats}, we consider the caveats and limitations to our main argument.  We conclude in Section~\ref{sec:discussion}. 
 
\section{Angular Momentum Evolution in a `Classic' Scenario}
\label{sec:a range}

The  \textit{classical scenario} for the merger of stellar-mass BHs from an isolated stellar field binary was worked out nearly three decades ago with subsequent modern improvements \citep{Phinney1991,Tutukov1993,Belczynski2016}. In this picture, prior to the collapse of the second BH, the progenitor system is an isolated stellar field binary that is composed of a Wolf--Rayet (WR) star with a mass $M$ and a BH with a mass $M/q$, $q$ being defined as the mass ratio.  

We now investigate the angular momentum evolution of the WR star. The final angular momentum of the WR star is the angular momentum of the second BH under the assumption that there is no mass-loss during the collapse. Mass-loss also increases the initial semi-major axis of the BH+BH orbit compared with the BH+WR orbit. We will revisit this assumption in Section~\ref{sec:discussion} and show that mass-loss during the collapse does not change our result significantly. We further assume that the kick velocity of the second BH at birth is very small compared with the orbital velocity.   Any scenario involving a large BH natal kick is ruled out by the existence of the merger.

For an initial orbital semi major axis, $d$, we can normalize the dimensionless spin of the star, $a$, to the orbital angular velocity, $\omega=\sqrt{G(M+M/q)/d^{3}}$ :
\begin{eqnarray}\label{eq:a}
a&=&\frac{cJ}{GM^{2}}=\frac{cr_{g}^{2} R^{2}}{GM}\left(\frac{1+q}{2q}\right)^{1/2}\left(\frac{2GM}{d^{3}}\right)^{1/2}\frac{\Omega}{\omega} \nonumber\\
&\equiv&\async\frac{\Omega}{\omega}.
\end{eqnarray}
For synchronization between the stellar spin and the orbit ($\Omega=\omega$), we define $a=\async$. 

Because the merger time due to gravitational wave emission, $t_{\rm merge}$, also depends simply on the initial orbital properties and component masses \citep{Peters1964}:
\begin{equation}\label{eq:tmerger}
\tmerge=\frac{5}{512}\frac{c^{5}}{G^{3}M^{3}}\frac{2q^{2}}{1+q}d^{4},
\end{equation}
we can write $\async$ as follows:
\begin{eqnarray}\label{eq:a2}
\async&\approx&0.44\left(\frac{q^{2}(1+q)}{2}\right)^{1/8}\left(\frac{r_{g}^{2}}{0.075}\right)\left(\frac{R}{2\,R_{\odot}}\right)^{2}\times \nonumber \\
&&\left(\frac{M}{30\,M_{\odot}}\right)^{-13/8}\left(\frac{\tmerge}{1\,\textrm{Gyr}}\right)^{-3/8}.
\end{eqnarray}
Although not used later on, it may be useful for subsequent studies to derive similar expression for the initial orbital period, equatorial velocity, and the break-up fraction:
\begin{eqnarray}\label{eq:P, v and f}
P&\approx&1.81\left(\frac{q^{2}(1+q)}{2}\right)^{-1/8}\left(\frac{M}{30\,M_{\odot}}\right)^{5/8}\left(\frac{\tmerge}{1\,\textrm{Gyr}}\right)^{3/8}\,\textrm{day},\nonumber\\
\veq&=&R\Omega=R\left(\frac{1+q}{2q}\right)^{1/2}\left(\frac{2GM}{d^{3}}\right)^{1/2}\frac{\Omega}{\omega}\nonumber\\
&\approx&55.8\left(\frac{q^{2}(1+q)}{2}\right)^{1/8}\left(\frac{R}{2R_{\odot}}\right)\times\nonumber\\
&&\left(\frac{M}{30\,M_{\odot}}\right)^{-5/8}\left(\frac{\tmerge}{1\,\textrm{Gyr}}\right)^{-3/8}\frac{\Omega}{\omega}\,\textrm{km}\,\textrm{s}^{-1},\nonumber\\
\fb&=&\frac{\Omega^{2}R^{3}}{GM}=2\left(\frac{R}{d}\right)^{3}\left(\frac{1+q}{2q}\right)\left(\frac{\Omega}{\omega}\right)^{2}\nonumber\\
&\approx&1.09\cdot10^{-3}\left(\frac{q^{2}(1+q)}{2}\right)^{1/4}\left(\frac{R}{2R_{\odot}}\right)^{3}\times\nonumber\\
&&\left(\frac{M}{30\,M_{\odot}}\right)^{-9/4}\left(\frac{\tmerge}{1\,\textrm{Gyr}}\right)^{-3/4}\left(\frac{\Omega}{\omega}\right)^{2}.
\end{eqnarray}

The angular momentum evolution of the WR star is determined by two competing process. Stellar winds decrease the angular momentum of the star (Section~\ref{sec:wind}), while torque applied to the star by the BH increases the angular momentum, driving it towards $\async$ (Section~\ref{sec:torque}).   We now investigate the impact of each of these processes on the total angular momentum evolution.

\subsection{Wind Losses}
\label{sec:wind}
Mass-loss from WR stars is complicated theoretically and observationally, thus the estimated rates are highly uncertain.  A typical estimate is $\approx10^{-5}\,M_{\odot}\,\textrm{yr}^{-1}$ \citep[see][and references therein]{Crowther2007}, while \citet{Nugis2000} estimate $M/\dot{M}\approx10^{6}\,\textrm{yr}$ independent of mass. An expected dependence of the mass-loss rate on metallicity further complicates the estimates \citep{Vink2001}. Here, we assume that over the time of the WR phase, $\twr$, the star does not lose a significant fraction of its mass, i.e. $\twr\simlt M/\dot{M}$. For massive $\simgt10\,M_{\odot}$ WR stars, this is a reasonable assumption, as the estimated lifetime is $\twr\approx3\cdot10^{5}\,\textrm{yr}$ \citep[see][for a recent study]{McClelland2016}, and $M/\dot{M}\simgt10^{6}\,\textrm{yr}$. We are interested in $\twind$, the time-scale for loss of angular momentum resulting from winds. For $\twind\simlt\twr$ the angular momentum is exponentially suppressed due to the wind. We now estimate this time-scale for the system of interest. 

Assuming the stellar radius is constant during the mass loss and that the mass is lost from a spherical shell at the stellar surface \citep[see discussion in][]{Ro2016}, we can write the rate of angular momentum loss as follows \citep{Packet1981}:  
\begin{eqnarray}\label{eq:wind 2}
\dot{J}&=&\dot{I}\Omega=\frac{2\dot{M}R^{2}\Omega}{3}.
\end{eqnarray}
We can estimate $\twind$ by taking the ratio between $J$ and $\dot{J}$, 
\begin{eqnarray}\label{eq:wind time}
\twind\equiv\frac{a}{\dot{a}}\approx\frac{J}{\dot{J}}\approx\frac{3}{2}r_{g}^{2}\frac{M}{\dot{M}}\approx0.1\left(\frac{r_{g}^{2}}{0.075}\right)\frac{M}{\dot{M}},
\end{eqnarray}
which suggests $\twind\simgt10^{5}\,\textrm{yr}$. In the case that the wind is magnetized, the specific angular momentum of the wind is larger than the previous estimate by a factor $R_{\textrm{A}}/R$, where $R_{\textrm{A}}$ is the Alfv\`en radius, which can significantly decrease the value of $\twind$.

We can use the observed rotational periods of isolated WR stars to estimate $\twind$. There are a few WR stars with a claimed detection of rotation. Two of them show no signs of a companion and quite high angular spin velocities; WR 134 with $\Omega\approx2.7\,\textrm{day}^{-1}$ \citep{McCandliss1994,Morel1999} and WR 6 with $\Omega\approx1.7\,\textrm{day}^{-1}$ \citep{St-Louis1995,Morel1997}. Assume that in the beginning of the WR phase ($t=0$), the break-up fraction at the surface was $f$, and that the star loses angular momentum only by the wind, then the angular spin velocity is given by
 \begin{eqnarray}\label{eq:omega evol}
\Omega(t)&=&\sqrt{\frac{G M f}{R^3}}\exp(-t/\twind)\nonumber\\
\Rightarrow&&\twind=\frac{t}{\ln\left(\sqrt{\frac{GMf}{R^{3}\Omega^{2}}}\right)}.
\end{eqnarray}
The factor in the denominator is 2--3 for the observed $\Omega$ values, typical $R$ and $M$ values, and $f=0.01$. Therefore, in the case that the age of WR 134 and WR 6 is some significant fraction of $\twr$, we can estimate $\twind\sim10^5\,\textrm{yr}$. Another possibility is that these stars are very young and this is the reason they are observed to be rotating fast. In this case $\twind$ is smaller by a factor of $t/\twr$.

\subsection{Torque Gains}
\label{sec:torque}
The torque applied to a star in a binary system was first calculated by \citet{Zahn1975}. We use the following expression for the torque applied to the star by the BH, as further described in \citet{Kushnir:2016b}\footnote{In our case we can assume that the normalized apparent frequency of the tide, $s=2|\omega-\Omega|\sqrt{R^{3}/GM}$, satisfies $s^{-1}\gg1$:
\begin{eqnarray}\label{eq:s limit}
s&<&2\omega\left(\frac{R^{3}}{GM}\right)^{1/2}=2^{3/2}\left(\frac{1+q}{2q}\right)^{1/2}\left(\frac{R}{d}\right)^{3/2}, \nonumber \\
\Rightarrow s^{-1}&\gtrsim&15.2\left(\frac{q^{2}(1+q)}{2}\right)^{-1/8}\left(\frac{R}{2R_{\odot}}\right)^{-3/2}\nonumber\\
&\times&\left(\frac{\tmerge}{\textrm{Gyr}}\right)^{3/8}\left(\frac{M}{30M_{\odot}}\right)^{9/8}.
\end{eqnarray}
In this case, the forced oscillations in the envelope behave like a purely travelling wave. For the case that $s^{-1}$ is small compared with unity, one must further consider the damping of the waves in the envelope, which is not the relevant regime and thus outside the scope of this paper.}:
\begin{equation}\label{eq:torque app}
\tau\approx\frac{G(M/q)^{2}}{r_{c}}\left(\frac{r_{c}}{d}\right)^{6}\left[\frac{4(\omega-\Omega)^{2}r_{c}^{3}}{GM_c}\right]^{4/3}\frac{\rho_c}{\bar{\rho}_c}\left(1-\frac{\rho_c}{\bar{\rho}_c}\right)^2,
\end{equation}
where $\rho$ is the density, $\bar{\rho}$ is the mean density inside the sphere of radius $r$, and the subscript $c$ indicates values at the convective core boundary. Given this expression for the torque, the change in $a$ can be computed:
\begin{equation}\label{eq:a due to torque}
\dot{a}=\frac{c}{GM^{2}}\tau\equiv\frac{\async}{\ttau}\left|1-\frac{a}{\async}\right|^{8/3},
\end{equation}
where the relevant time-scale is
\begin{eqnarray}\label{eq:torque time}
\ttau&=&\async\frac{GM^{2}}{c\tau\left(\Omega=0\right)}\nonumber\\
&=&2^{-21/6}\left(\frac{512}{5}\right)^{17/8}q^{-1/8}\left(\frac{1+q}{2q}\right)^{31/24}r_{g}^{2}\left(\frac{R}{r_{c}}\right)^{2}\nonumber\\
&\times&\left(\frac{GM}{r_c c^{2}}\right)^{109/24}\left(\frac{GM_{c}}{r_c c^2}\right)^{4/3}\left(\frac{\rho_c}{\bar{\rho}_c}\right)^{-1}\left(1-\frac{\rho_c}{\bar{\rho}_c}\right)^{-2}\nonumber\\
&\times&\left(\frac{c\tmerge}{r_c}\right)^{9/8}\tmerge\nonumber\\
&\equiv&q^{-1/8}\left(\frac{1+q}{2q}\right)^{31/24}\ftau(\tmerge)\tmerge
\end{eqnarray}

In order to estimate this time-scale, we require knowledge of the physical structure ($R$, $r_c$, $M_c$, $r_g^2$) of the WR star. The time-scale is especially sensitive to $r_c$ as it scales with $r_c^{-9}$, and much less sensitive to the other parameters. Note that $\ttau$ scales as $\tmerge^{17/8}$, and since we are interested in estimating $\tmerge$ to an order of a magnitude, it is sufficient to estimate $\ftau$ (at some fixed $\tmerge$) to factor of a few, which we presently evaluate.

\subsection{WR stellar models}
\label{sec:WR}

The values of $R$, $r_{c}$, $M_c$, and $r_{g}^{2}$ are challenging to directly observe, due to optically thick stellar winds that obscure the stellar surface, making the interpretation of measurements challenging. It is therefore uncertain to use empirical estimates of these quantities, and we are forced to rely on stellar evolution models to obtain estimates of these parameters.  

We construct stellar evolution models using the publicly available package MESA version 6596 \citep{2011ApJS..192....3P,2013ApJS..208....4P,2015ApJS..220...15P}. We aim to cover a wide range of masses during the WR phase ($\approx[5,32] M_{\odot}$), and for that we select from a set of models having zero-age main sequence masses in the range $[40,100] M_{\odot}$, metallicities between $[0.01,1] Z_{\odot}$, and initial rotation between $[0.4,0.6]$ of breakup. We use the WR model profiles during the epoch where the time to core-collapse is greater than $10^4\,\textrm{yr}$ and the stellar radius is smaller than $2 R_{\textrm{SM}}$, where $R_{\textrm{SM}}$ is the WR radius according to \citet{Schaerer1992}.

In all models, mass-loss was determined according to the `Dutch' recipe in MESA, combining the rates from \citet{2009A&A...497..255G, 1990A&A...231..134N, Nugis2000}, and \citet{Vink2001}, with a coefficient $\eta=1$, convection according to the Ledoux criterion, with mixing length parameter $\alpha_{\textrm{mlt}}=2$, semi-convection efficiency parameter $\alpha_{\textrm{sc}}=0.1$ \citep[equation 12]{2013ApJS..208....4P}, and exponential overshoot parameter $f=0.008$ \citep[equation 2]{2011ApJS..192....3P}. For the atmosphere boundary condition we use the simple option of MESA \citep[eq. 3]{2011ApJS..192....3P}.
  
\begin{figure}
\includegraphics[width=0.5\textwidth]{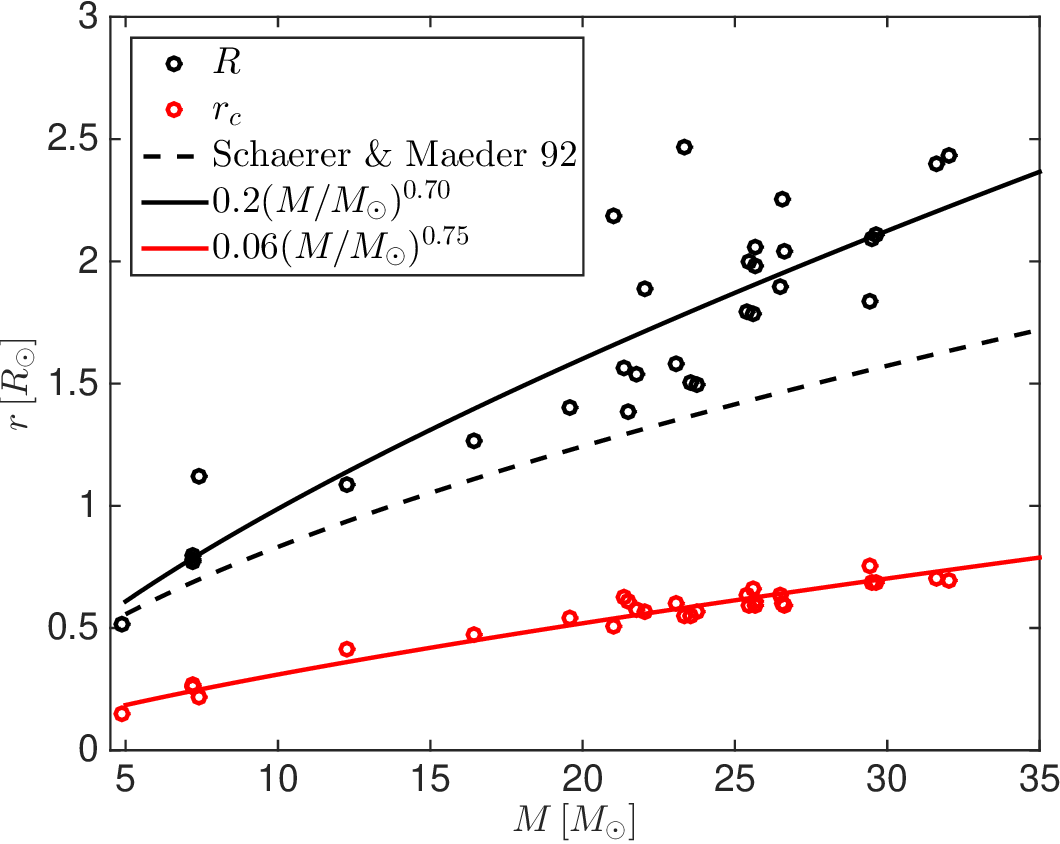}
\caption{$R$ (black curves) and $r_c$ (red curves) for a series of stellar models. The open symbols show our results for a series of models computed by MESA.  Simple fits to these points are shown by the solid black and red curves.  The dashed black curve shows the estimate of $R$ from \citet{Schaerer1992} based on the stellar evolution models of \citet{Schaller1992}. The predicted larger radii ($\sim30\%$) at fixed mass in our models relative to the \citet{Schaller1992} models, likely due to our simplified treatment of the stellar atmosphere, do not significantly impact our final results. 
\label{fig:WRBasic}}
\end{figure}

In Figure~\ref{fig:WRBasic}, we present $R$ and $r_{c}$ as a function of $M$ for the models we consider.  Our results can be fit as $\log_{10}(R/R_{\odot})\approx-0.70+0.70\log_{10}(M/M_{\odot})$ and as $\log_{10}(r_{c}/R_{\odot})\approx-1.25+0.75\log_{10}(M/M_{\odot})$. The profiles are well described by polytropes with $n\approx2.5$--$3.5$ and $r_{g}^{2}$ is in the range $\approx0.05$--$0.09$, which roughly corresponds to these polytropes. Our calculated stellar radii are compared with the stellar evolution models of \citet{Schaller1992}, evaluated in \citet{Schaerer1992}. We predict larger radii ($\sim30\%$) at fixed mass in our models relative to the \citet{Schaller1992}, possibly because of the simple MESA photosphere boundary condition that we used \citep[see discussion in][]{Ro2016}. We note that generally, stellar evolution models predict radii that are significantly smaller than those derived from atmospheric models \citep[$\approx3\,R_{\odot}$ in some cases; see][and references therein]{Crowther2007}, and therefore should be considered at the moment as a rough estimate. Nevertheless, the uncertainty in the stellar radii is insignificant for our final results, which depend much more strongly on the core radii. 
 
\begin{figure}
\includegraphics[width=0.5\textwidth]{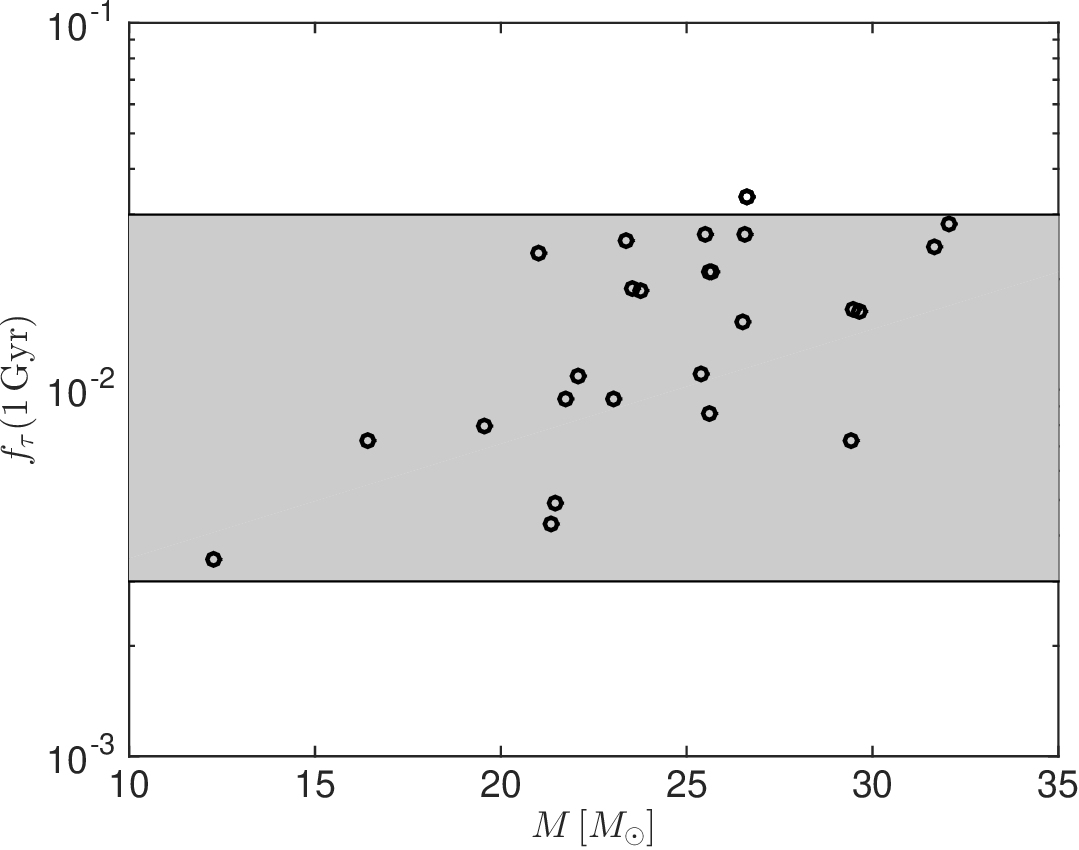}
\caption{The pre-factor from Equation~\eqref{eq:torque time}, $\ftau(1\,\textrm{Gyr})$, as function of the stellar mass. The main uncertainty in the value of $\ftau$ is the core radius $r_c$, as $\ftau\propto r_c^{-9}$. The scatter of $\ftau$ at fixed mass is completely driven by the scatter of $r_c$ in Figure~\ref{fig:WRBasic}. We approximate $\ftau(1\,\textrm{Gyr})\approx0.01$, which reproduces the numerical results to better than a factor of $3$, as demonstrated by the shaded band.
\label{fig:WRfactau}}
\end{figure}

Instead of fitting separately for each parameter, we directly calculate, for each WR model, $\ftau(1\,\textrm{Gyr})$ from Equation~\eqref{eq:torque time}. The results are presented in Figure~\ref{fig:WRfactau}. The main uncertainty in the value of $\ftau$ is the core radius $r_c$, as $\ftau\propto r_c^{-9}$. The scatter of $\ftau$ at fixed mass is completely driven by the scatter of $r_c$ in Figure~\ref{fig:WRBasic}. We approximate $\ftau(1\,\textrm{Gyr})\approx0.01$, which reproduces the numerical results to better than a factor of $3$, as demonstrated by the shaded band in Figure~\ref{fig:WRfactau}. This allows us to rewrite Equation~\eqref{eq:torque time} simply as
\begin{eqnarray}\label{eq:torque time2}
\ttau\approx10^{7}q^{-1/8}\left(\frac{1+q}{2q}\right)^{31/24}\left(\frac{\tmerge}{1\,\textrm{Gyr}}\right)^{17/8}\,\textrm{yr}.
\end{eqnarray}
We are now able to evaluate the angular momentum evolution of the system prior to merger.
 
\section{Implications}
\label{sec:implications}

At the beginning of the WR phase, we expect $0\le a\le \async$. That is, following the previous evolution phases of the system, we neither expect the star to be spinning faster than the synchronization value nor retrograde. We can therefore compute the evolution of the spin taking into account the breaking and torquing described above. Including both effects, the evolution of $a$ is simply
\begin{equation}\label{eq:a evolution}
\dot{a}=\frac{\async}{\ttau}\left(1-\frac{a}{\async}\right)^{8/3}-\frac{a}{\twind}.
\end{equation}

We can integrate Equation~\eqref{eq:a evolution} directly to find $a(\twr)$ given $\twr$, $\twind$, $\ttau$, and $a(0)$. The angular momentum of the second BH is given by $\min(a(\twr),1)$. We can gain some insight into the solution of Equation~\eqref{eq:a evolution} by writing it as   
\begin{equation}\label{eq:x evolution}
x'=\frac{\twind}{\ttau}\left(1-x\right)^{8/3}-x,
\end{equation}
where $x=a/\async$, prime indicated a derivative with respect to $t/\twind$, and $0\le x(0)\le1$. For large enough $\twr/\twind$ we have $x(\twr/\twind)\approx \xeq$, where 
\begin{equation}\label{eq:xeq}
\frac{\twind}{\ttau}\left(1-\xeq\right)^{8/3}=\xeq.
\end{equation} 
The solution of Equation~\eqref{eq:x evolution} in a few representative cases can be approximated as follows.
\begin{enumerate}
\item $\twind\simgt\twr$, $x(0)=1$; $x(\twr/\twind)\approx\max(1-\twr/\twind,\xeq)$.
\item $\twind\simgt\twr$, $x(0)=0$; $x(\twr/\twind)\approx\min(\twr/\ttau,\xeq)$. 
\item $\twind\simlt\twr$, $x(0)=1$;  $x(\twr/\twind)\approx\max[\exp(-(\twr/\twind),\xeq]$. 
\item $\twind\simlt\twr$, $x(0)=0$;  $x(\twr/\twind)\approx\xeq$. 
\end{enumerate}

\begin{figure}
             \includegraphics[width=0.23\textwidth]{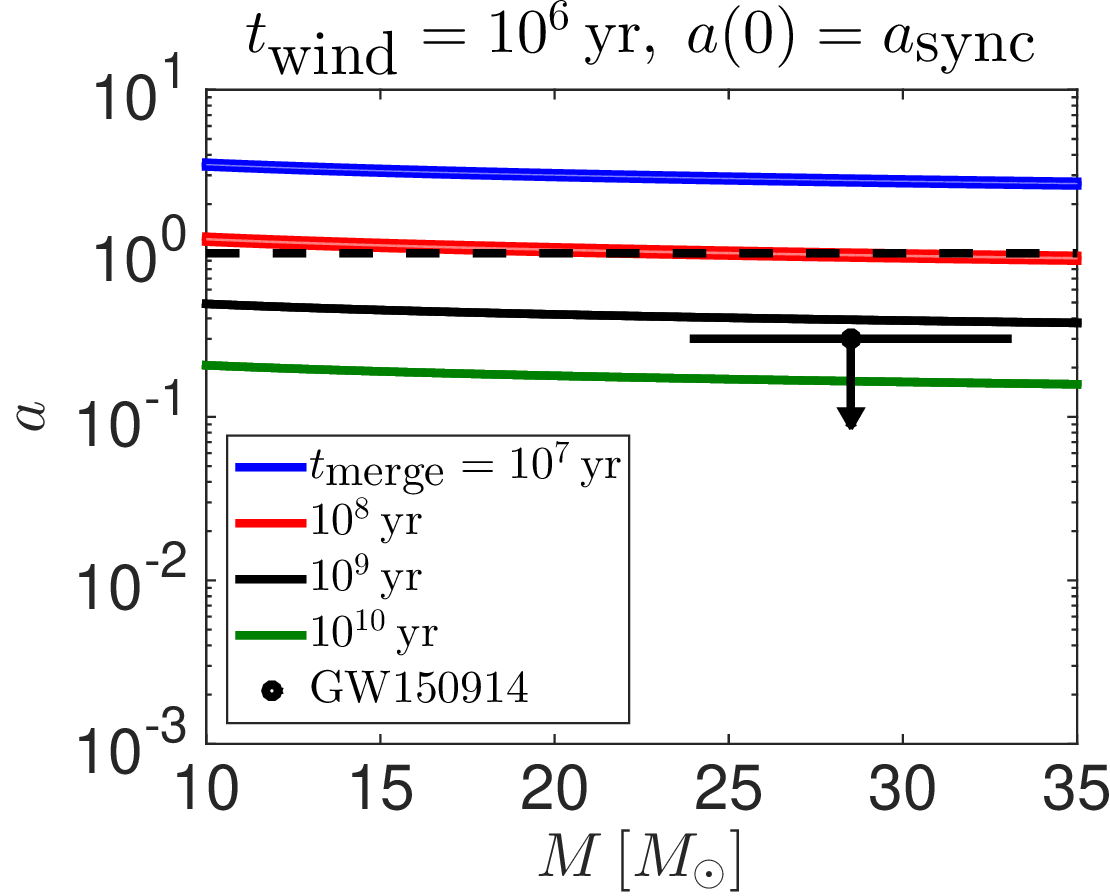}
             \includegraphics[width=0.23\textwidth]{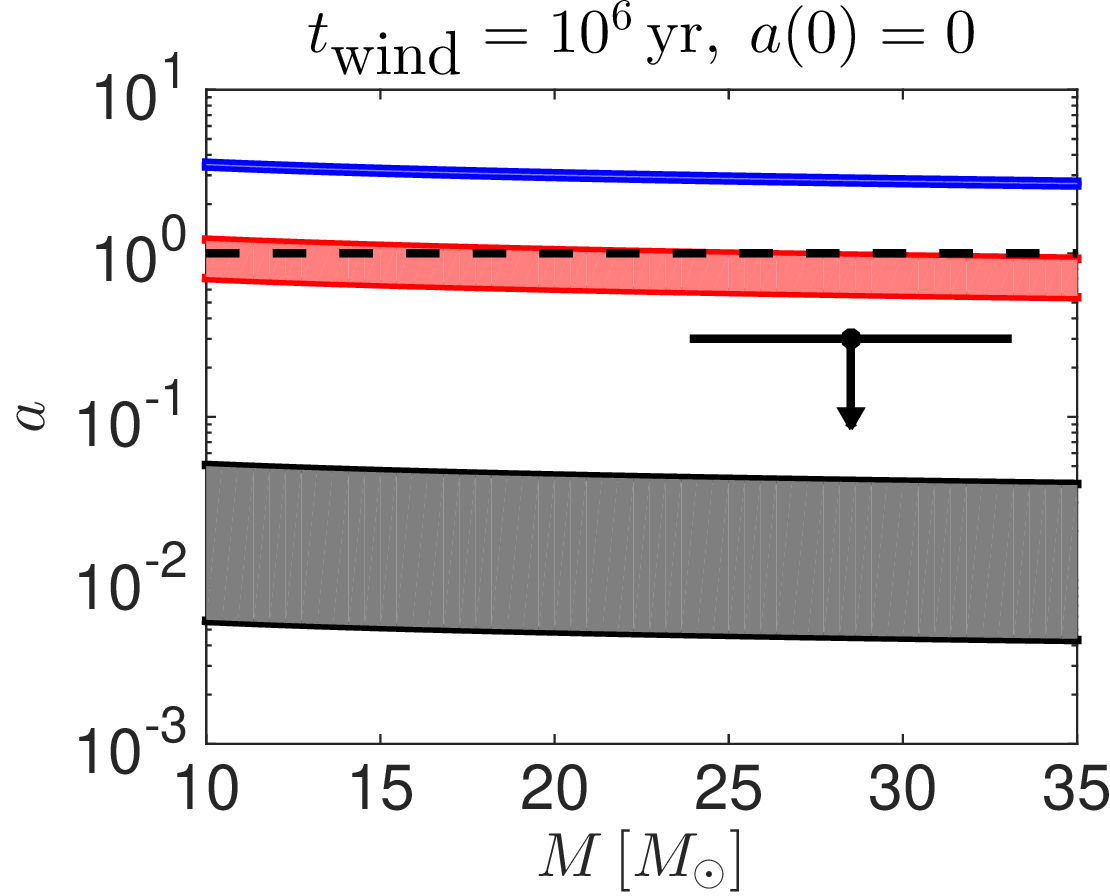}
             \includegraphics[width=0.23\textwidth]{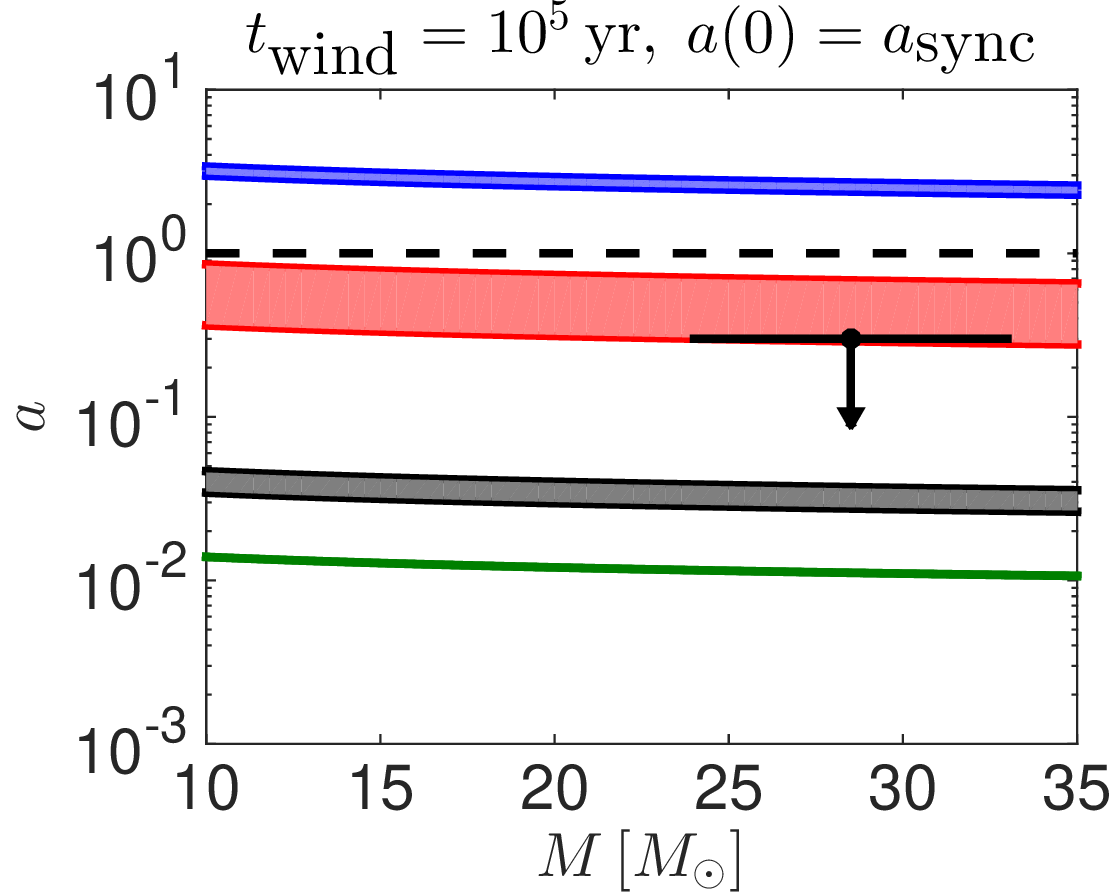}
             \includegraphics[width=0.23\textwidth]{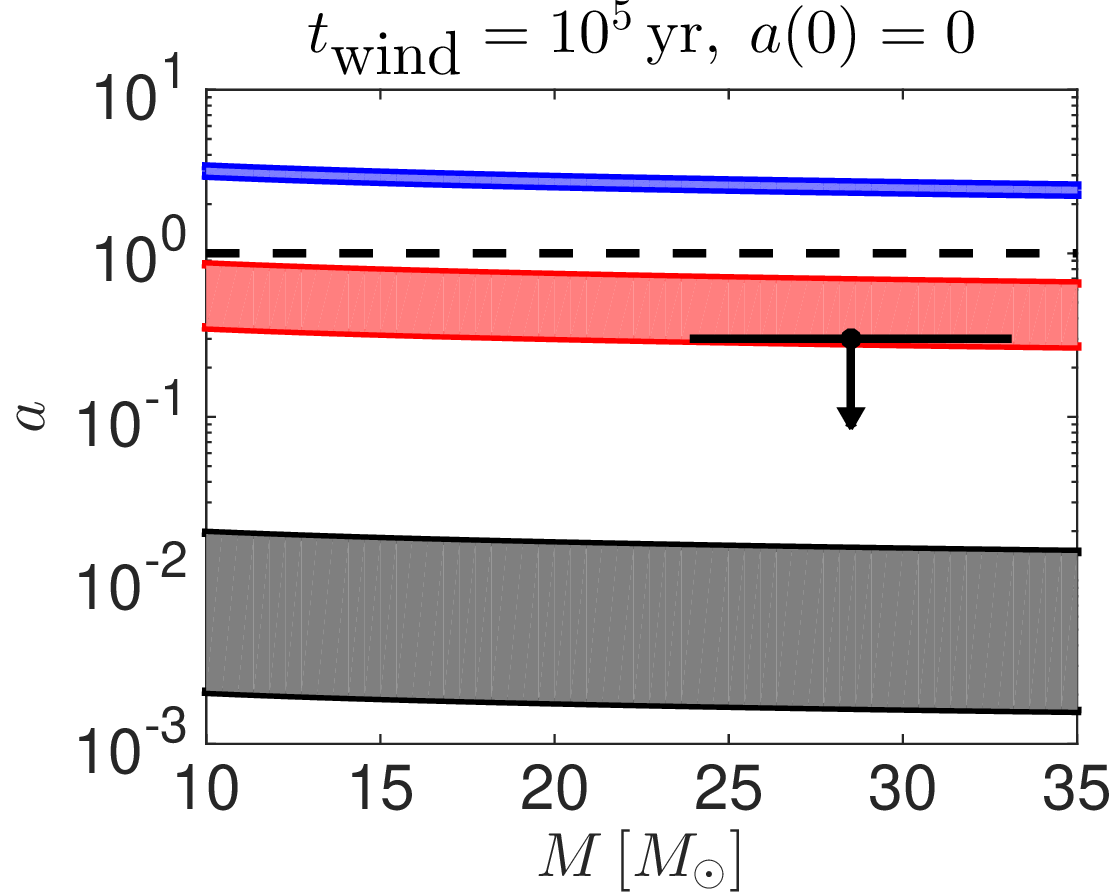}
      \caption{(a)--(d) Dimensionless spin-parameter of the secondary BH as a function of $M$ for different limiting cases.  The black symbol shows $\gw$. Blue, red, black, and green lines show merger times of $10^7$, $10^8$, $10^9$, and $10^{10}\,\textrm{yr}$. The shaded band for each merger time represent the range $3\cdot10^{-3}\le\ftau(1\,\textrm{Gyr})\le3\cdot10^{-2}$, which corresponds to the uncertainty in its value (shaded band in Figure~\ref{fig:WRfactau}). Lower values of $a$ correspond to higher values of $\ftau$. We choose $\twr/\twind=0.3$ (upper panels) and $\twr/\twind=3$ (lower panels) to span reasonable parameters for the wind. We study cases when the initial spin is close to zero (right-hand panels) or close to $\async$ (left-hand panels). For convenience, we plot values above the $a=1$ limit (dashed line).}
   \label{fig:aM}
\end{figure}

We solve Equation~\eqref{eq:a evolution} for four representative examples and show the results in Figure~\ref{fig:aM}. We consider cases when $\twind\simgt\twr$ or $\twind\simlt\twr$ and when the initial spin is close to zero or close to $\async$. In our examples for Figure~\ref{fig:aM} we choose $\twr/\twind=0.3$ ($\twind=10^6\,\textrm{yr}$) and $\twr/\twind=3$ ($\twind=10^5\,\textrm{yr}$) so as to span reasonable parameters for the wind. We calculated for both $\ftau(1\,\textrm{Gyr})=3\cdot10^{-3}$ and $\ftau(1\,\textrm{Gyr})=3\cdot10^{-2}$, to span the uncertainty in its value. We further assume that $q=1$, consistent with \gw. Our results are compared with the limit from $\gw$, $a<0.3$ at $90\%$ probability \citep{LIGO2016}.  In our comparison, we use the values for the less massive BH as the more massive BH yields slightly stronger constraints. We conclude that in order for $a$ not to exceed the observed limit, we must have $\tmerge\simgt10^{8}\,\textrm{yr}$. This lower limit is driven by the cases with small $\twind$, and is independent of the value of $a(0)$. Smaller values of $\twind$ would decrease the derived lower limit. For $\twind=10^{4}\,\textrm{yr}$, we obtained $\tmerge\simgt\textrm{few}\times10^{7}\,\textrm{yr}$. Models of the classical scenario can be compared with the derived lower limit of $\tmerge$. For example, the vast majority of models in \citet{Belczynski2016} fulfil this condition. Note that the scenario of no winds and $a(0)=\async$ is not far from being ruled out, as using the age of the Universe for $\tmerge$ in Equation~\eqref{eq:a} leads in this case to $a=\async\approx0.15$.

\section{Caveats}
\label{sec:caveats}

\begin{figure}
\includegraphics[width=0.5\textwidth]{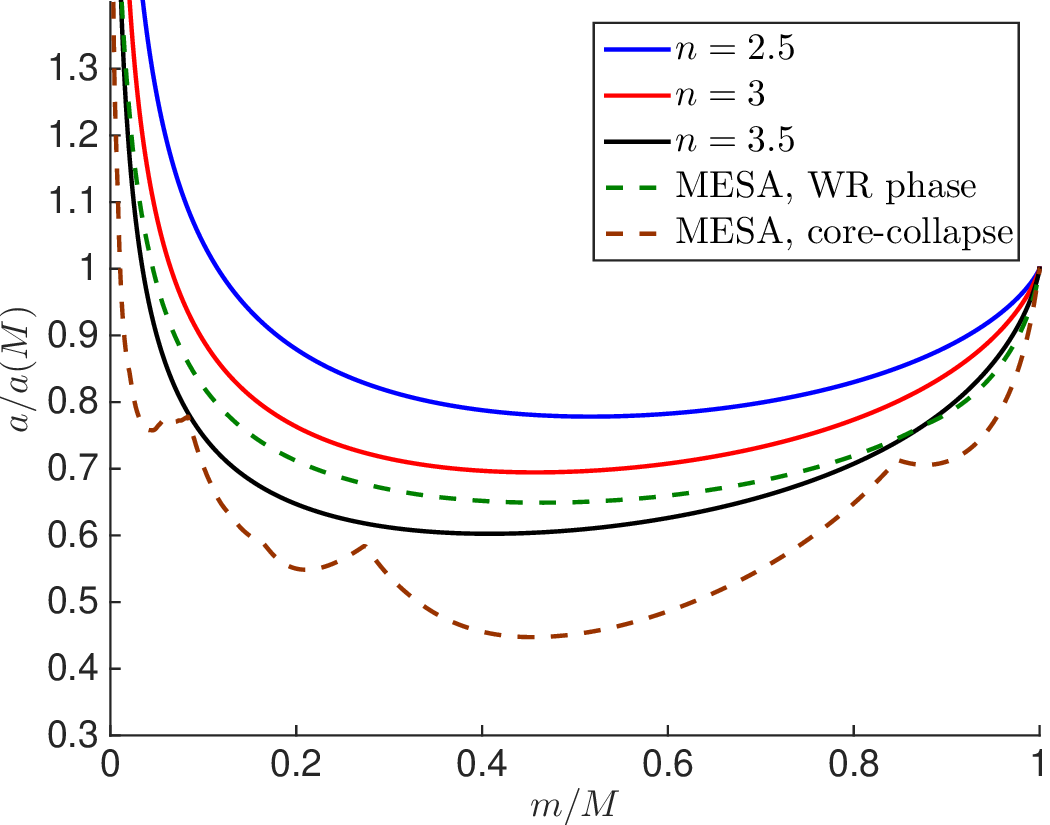}
\caption{ Effect of mass-loss on angular momentum evolution. We compute the change in $a$ for three different polytropes in solid-body rotation as mass is changed relative to an initial mass (solid). We compare the polytrope estimate to one of our MESA models with a WR mass of $M\approx30\,M_{\odot}$ for both the WR phase prior to the last burning stages (dashed green) and for the time of core-collapse (dashed brown), which demonstrate the robustness of the polytrope estimates. The possible decrease in $a$ is modest ($20$--$50\%$), and obtained for substantial mass-loss.
\label{fig:apoly}}
\end{figure}

We assume there is no mass-loss during the collapse. Mass-loss changes our analysis in two ways. First, mass-loss widens the initial separation of the BH+BH orbit compared with the BH+WR orbit. This effect makes the merger time longer, which does not affect the derived lower limit for $\tmerge$. Secondly, angular momentum is lost with the mass ejection, which may change our estimate for the spin evolution of the BH. We estimate the possible decrease in $a$ by considering in Figure~\ref{fig:apoly} polytropes with solid body rotation, which is the expected configuration during the WR phase. The very last burning stages of the WR star prior to collapse are expected to be sufficiently fast, such that angular momentum transfer can be neglected and each mass shell retains its original $a$ value. We compare the polytrope estimate to one of our MESA models with a WR mass of $M\approx30\,M_{\odot}$ for both the WR phase prior to the last burning stages and for the time of core-collapse, which demonstrate the robustness of the polytrope estimates. We show in Figure~\ref{fig:apoly} that the value of $a$ can decrease by $20$--$50\%$ for a large mass ejection, which does not significantly affect our analysis. Moreover, if indeed the collapse was accompanied by a supernova, then it should be a Type Ibc supernova, for which the mass of the ejecta is small compared to $30\,M_{\odot}$ \citep[e.g.,][]{Lyman2014}. Indeed, such a system is an expected outcome of a Type Ibc supernova for collapse-induced thermonuclear explosions \citep{Kushnir:2015mca,Kushnir:2015vka}. 

We thus conclude that our analysis is robust to the simplifications we have made.

\section{Discussion}
\label{sec:discussion}

We have provided a simple framework for using the observed (low) BH spin of \gw to constrain the BH merger time and thus the characteristics of the progenitor system. We have shown that already with a single event, the observed constraint on $\chieff$ comes astonishingly close to ruling out certain scenarios. Future events will allow us to map the true delay-time distribution as a function of progenitor mass -- a critical quantity in understanding the latest-stage evolution of the progenitors --- for comparison with predictions. We eagerly await the results of future observations of gravity waves from these truly remarkable events and expect that the measurement of the spins will be very informative to constrain the different formation channels that have been proposed.


\section*{Acknowledgements}
We thank Ben Bar-Or, Jeremy Goodman, Boaz Katz, Roman Rafikov, and Eli Waxman for discussions. MZ is
supported in part by the NSF grants PHY-1213563, AST-1409709, and PHY-1521097.   JAK gratefully acknowledges support from the Institute for Advanced Study.





\begin{thebibliography}{99}
\bibitem[Abbott et al.(2016)]{Abbott2016PhRvL} Abbott, B.~P., Abbott, R., Abbott, T.~D., et al.\ 2016, Physical Review Letters, 116, 061102
\bibitem[Abbott et al.(2016)]{Abbott2016ApJL} Abbott, B.~P., Abbott, R., Abbott, T.~D., et al.\ 2016, \apjl, 818, L22
\bibitem[Abbott et al.(2016)]{LIGO2016} Abbott, B.~P., Abbott, R., Abbott, T.~D., et al.\ 2016, Physical Review Letters, 116, 241102 
 \bibitem[Belczynski et al.(2016)]{Belczynski2016} Belczynski, K., Holz, D.~E., Bulik, T., \& O'Shaughnessy, R.\ 2016, \nat, 534, 512 
\bibitem[Crowther(2007)]{Crowther2007} Crowther, P.~A.\ 2007, \araa, 45, 177 
\bibitem[de Mink \& Mandel(2016)]{dMink2016} de Mink, S.~E., \& Mandel, I.\ 2016, \mnras, 460, 3545 
\bibitem[Fryer et al.(2001)]{Fryer2001} Fryer, C.~L., Woosley, S.~E., \& Heger, A.\ 2001, \apj, 550, 372 
\bibitem[Glebbeek et 
al.(2009)]{2009A&A...497..255G} Glebbeek, E., Gaburov, E., de Mink, S.~E., Pols, O.~R., \& Portegies Zwart, S.~F.\ 2009, \aap, 497, 255 
\bibitem[{Kushnir(2015{\natexlab{a}})}]{Kushnir:2015mca} Kushnir, D.\ 2015{\natexlab{a}}, arXiv:1502.03111 
\bibitem[{Kushnir(2015{\natexlab{b}})}]{Kushnir:2015vka} Kushnir, D.\ 2015{\natexlab{b}}, 
arXiv:1506.02655 
\bibitem[{Kushnir(2016{\natexlab{b}})}]{Kushnir:2016b} Kushnir, D., Zaldarriaga, M., Kollmeier, J.~A., \& Waldman, R.\ 2016, arXiv:1605.03810 
\bibitem[Loeb(2016)]{Loeb2016} Loeb, A.\ 2016, \apjl, 819, L21
\bibitem[Lyman et al.(2016)]{Lyman2014} Lyman, J.~D., Bersier, D., James, P.~A., et al.\ 2016, \mnras, 457, 328
\bibitem[Mandel \& de Mink(2016)]{Mandel2016} Mandel, I., \& de Mink, S.~E.\ 2016, \mnras, 458, 2634
\bibitem[Marchant et al.(2016)]{Marchant2016} Marchant, P., Langer, N., Podsiadlowski, P., Tauris, T.~M., \& Moriya, T.~J.\ 2016, \aap, 588, A50
\bibitem[McCandliss et al.(1994)]{McCandliss1994} McCandliss, S.~R., Bohannan, B., Robert, C., \& Moffat, A.~F.~J.\ 1994, \apss, 221, 155 
\bibitem[McClelland \& Eldridge(2016)]{2McClelland2016} McClelland, L.~A.~S., \& Eldridge, J.~J.\ 2016, \mnras, 459, 1505 
\bibitem[Morel et al.(1997)]{Morel1997} Morel, T., St-Louis, N., \& Marchenko, S.~V.\ 1997, \apj, 482, 470 
\bibitem[Morel et al.(1999)]{Morel1999} Morel, T., Marchenko, S.~V., Eenens, P.~R.~J., et al.\ 1999, \apj, 518, 428 
\bibitem[Nieuwenhuijzen 
\& de Jager(1990)]{1990A&A...231..134N} Nieuwenhuijzen, H., \& de Jager, C.\ 1990, \aap, 231, 134
\bibitem[Nugis \& Lamers(2000)]{Nugis2000} Nugis, T., \& Lamers, H.~J.~G.~L.~M.\ 2000, \aap, 360, 227
\bibitem[Packet(1981)]{Packet1981} Packet, W.\ 1981, \aap, 102, 17 
\bibitem[Paxton et al.(2011)]{2011ApJS..192....3P} Paxton, B., Bildsten,
L., Dotter, A., et al.\ 2011, \apjs, 192, 3
\bibitem[Paxton et al.(2013)]{2013ApJS..208....4P} Paxton, B., Cantiello,
M., Arras, P., et al.\ 2013, \apjs, 208, 4
\bibitem[Paxton et al.(2015)]{2015ApJS..220...15P} Paxton, B., Marchant,
P., Schwab, J., et al.\ 2015, \apjs, 220, 15
\bibitem[Peters(1964)]{Peters1964} Peters, P.~C.\ 1964, Physical Review, 136, 1224 
\bibitem[Perna et al.(2016)]{Perna2016} Perna, R., Lazzati, D., \& Giacomazzo, B.\ 2016, \apjl, 821, L18 
\bibitem[Phinney(1991)]{Phinney1991} Phinney, E.~S.\ 1991, \apjl, 380, L17 
\bibitem[Reisswig et al.(2013)]{Reisswig2013} Reisswig, C., Ott, C.~D., Abdikamalov, E., et al.\ 2013, Physical Review Letters, 111, 151101
\bibitem[Ro \& Matzner(2016)]{Ro2016} Ro, S., \& Matzner, C.~D.\ 2016, \apj, 821, 109 
\bibitem[Schaerer \& Maeder(1992)]{Schaerer1992} Schaerer, D., \& Maeder, A.\ 1992, \aap, 263, 129 
\bibitem[Schaller et al.(1992)]{Schaller1992} Schaller, G., Schaerer, D., Meynet, G., \& Maeder, A.\ 1992, \aaps, 96, 269
\bibitem[Sigurdsson \& Hernquist(1993)]{Sigurdsson1993} Sigurdsson, S., \& Hernquist, L.\ 1993, \nat, 364, 423 
\bibitem[St-Louis et al.(1995)]{St-Louis1995} St-Louis, N., Dalton, M.~J., Marchenko, S.~V., Moffat, A.~F.~J., \& Willis, A.~J.\ 1995, \apjl, 452, L57
\bibitem[Tutukov \& Yungelson(1993)]{Tutukov1993} Tutukov, A.~V., \& Yungelson, L.~R.\ 1993, \mnras, 260, 675 
\bibitem[Vink et al.(2001)]{Vink2001} Vink, J.~S., de Koter, A., \& Lamers, H.~J.~G.~L.~M.\ 2001, \aap, 369, 574 
\bibitem[Woosley(2016)]{Woosley2016} Woosley, S.~E.\ 2016, \apjl, 824, L10
\bibitem[Yoon et al.(2006)]{Yoon2006} Yoon, S.-C., Langer, N., \& Norman, C.\ 2006, \aap, 460, 199
\bibitem[Zahn(1975)]{Zahn1975} Zahn, J.-P.\ 1975, \aap, 41, 329
\end{thebibliography}





\bsp	
\label{lastpage}
\end{document}